# How Far Can Client-Only Solutions Go for Mobile Browser Speed?

Technical Report TR1215-2011, Rice University and Texas Instruments
[1]Zhen Wang, [2]Felix Xiaozhu Lin, [1,2]Lin Zhong, and [3]Mansoor Chishtie
[1]Dept. of ECE and [2]Dept of CS, Rice University, Houston, TX 77005     [3]Texas Instruments, Dallas, TX

## ABSTRACT

Mobile browser is known to be slow because of the bottleneck in resource loading. Client-only solutions to improve resource loading are attractive because they are immediately deployable, scalable, and secure. We present the first publicly known treatment of client-only solutions to understand how much they can improve mobile browser speed without infrastructure support. Leveraging an unprecedented set of web usage data collected from 24 iPhone users continuously over one year, we examine the three fundamental, orthogonal approaches a client-only solution can take: caching, prefetching, and speculative loading, which is first proposed and studied in this work. Speculative loading predicts and speculatively loads the subresources needed to open a web page once its URL is given. We show that while caching and prefetching are highly limited for mobile browsing, speculative loading can be significantly more effective. Empirically, we show that client-only solutions can improve the browser speed by about 1.4 second on average for web sites visited by the 24 iPhone users. We also report the design, realization, and evaluation of speculative loading in a WebKit-based browser called Tempo. On average, Tempo can reduce browser delay by 1 second (~20%).

## 1. Introduction

Web browser is one of the most important applications on smartphones but is known to be slow, taking many seconds to open a web page. The long delay harms mobile user experience and eventually discourages web-based business. For example, Google will lose up to 20% traffic with 500ms extra delay and Amazon will lose 1% sales with 100ms extra delay [1].

As shown by our previous work [2], the key to improve mobile browser is to speed up resource loading, the process that fetches the resources required to open a web page. Many effective solutions toward this end require infrastructure support, e.g., thin-client approaches [3-6], session-level techniques [7], prefetching [8-11] and SPDY, a new protocol [12]. They are limited in one or more of the following ways. First, solutions requiring web server support are difficult to deploy and may not work for legacy web sites. The adoption of a new protocol like SPDY [12] will take a long time, if it ever happens. Second, infrastructure support depends on server or proxy capabilities and do not scale up very well with the number of clients. For example, the failure of Amazon Web Services' cloud-computing infrastructure [13] took many websites down. Finally, solutions based on proxy support violate end-to-end security, which is crucial to secure websites.

Not surprisingly, solutions that do not rely on infrastructure support, or *client-only* solutions, are particularly attractive because they are immediately deployable, scalable, and secure. While client-only solutions are likely to be less effective than those leveraging infrastructure supports, it has been an open question how effective client-only solutions can be for mobile browsers. The challenge to answering this question has been the lack of data regarding the browsing behavior of mobile users.

The technical goal of this work is to answer this question, with the help of an unprecedented dataset of web browsing data continuously collected from 24 iPhone users over one year, or LiveLab traces [14]. In achieving our goal, we make four contributions. Firstly, we study browsing behavior of smartphone users and the web pages visited by them. We find that subresources needed for rendering a web page can be much more predictable than which webpage a user will visit because subresources have much higher revisit rate and a lot of them are shared by webpages from the same site.

Secondly, we quantitatively evaluate two popular client-only approaches: *caching* and *prefetching*. *Caching* seeks to store frequently used web resources locally, but we find that it has very limited effectiveness from the LiveLab traces: 60% of the requested resources are either expired or not in the cache. *Web prefetching*, e.g., [9-11], seeks to predict which webpage is likely to be visited by the user, and then fetches all the resources needed to render the page beforehand. While web prefetching with infrastructure support, e.g., [9-11], is known to be effective by aggregating many users' behavior, we find that, on smartphones, client-only prefetching is harmful and ineffective because web pages visited by mobile users are less predictable: over 75% of the visits in the LiveLab traces are to web pages visited only once.

Thirdly, we propose and study a new, orthogonal client-only approach: speculative loading. Given a web URL, speculative loading leverages concurrent connections available to modern browsers and loads subresources that are likely to be needed, concurrently with the main HTML file. To determine which subresources to load, the browser maps out how a website organizes resources based on the browsing history. We implement speculative loading in a WebKit-based browser called *Tempo* and evaluate it on real smartphones with 3G network. Evaluation shows that, on average, Tempo can improve browser speed by 1 second (~20%) with small data overhead. This will not only make web browsing noticeably faster but also may increase traffic to Google by up to 40% and increase Amazon sales by up to 10% according to [1].

Finally, because caching, prefetching, and speculative loading represent the three fundamental approaches that a client can improve resource loading in mobile browser, our findings enable us to answer the title question empirically: the upper bound of browser delay reduction from client-only solutions is about 1.4 second on average for the web sites visited by the LiveLab iPhone users. The client-only solutions are limited for four reasons: (*i*) a large portion of web resources are either not in the cache or their cached copies quickly expire; (*ii*) mobile browsing behaviors are not very predictable; (*iii*) a client cannot completely predict what resources are needed for a web page based on its user's history; (*iv*) the request-response model of HTTP [15] requires at least one request for each resource needed, which magnifies the impact of the relative long RTT of cellular networks. While 1.4 second is nontrivial, to make mobile browser instantly fast, infrastructure support is still necessary.

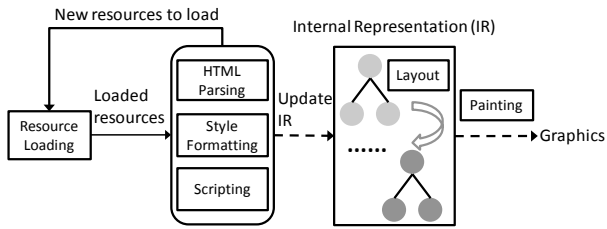

**Figure 1: The procedure of opening a webpage**

What Tempo achieves is very close to the upper bound. Tempo can also be combined with infrastructure support by providing the client knowledge of the server resources. For example, Tempo can help SPDY [12] to solve the race condition problem.

The rest of the paper is organized as follows. Section 2 introduces background and related work. Section 3 provides results from our characterization of mobile browsing and web pages. Section 4 investigates the three fundamental approaches available to client-only solutions. It provides an empirical analysis of the upper bound of improvement possible by client-only solutions. Section 5 presents the design and implementation of Tempo. Section 6 offers the results from lab and field based evaluation of Tempo. Section 7 concludes the paper.

## 2. Background and Related Work

We first provide an overview about how a mobile browser works using WebKit-based browsers. As illustrated by Figure 1, the procedure of opening a page involves six major operations that can be dynamically scheduled and concurrently executed. *Resource loading* fetches a resource given its URL, either from the remote web server or local cache. *HTMLParsing* (or *Parsing*), *StyleFormatting* (or *Style*) and *Scripting* process HTML documents, style constraints, e.g., Cascading Style Sheets (CSS), and JavaScript, respectively, and attach results to the internal representation (IR). *Layout* computes and updates the screen locations based on the recently updated IR. *Painting* employs the IR to generate the final graphical representation of the web page. It is important to note that these six operations do not form a simple pipeline in opening a page.

A browser usually needs multiple resources to open a webpage. A *resource* is an individual unit of content or code, usually uniquely identified by a web URL. The *main resource* is the first resource requested by the browser, usually an HTML document. After parsing the main resource, the browser may discover and load more resources that format, manipulate or provide additional content to the webpage. These resources, called *subresources*, usually correspond to CSS, JavaScript and picture files.

### 2.1 Why are Mobile Browsers Slow?

Browsers are well-known to be slow on mobile devices, taking many seconds to open a page, especially when using a cellular network. While prior work [16-18] suggests that several compute-intensive operations (*Style*, *Layout* and *Scripting*) should be the focus of optimization for browsers on PC, we recently showed [2] that the bottleneck of mobile browser performance is actually in *resource loading* due to long round trip time (RTT) and large number of total round trips. The RTT of typical 3G network is around 200ms [19], much longer than that of Ethernet network, and improves in a much slower pace than bandwidth. Moreover, resource loading in existing browsers is not fully parallel, resulting in a large number of round trips. Especially, subresources can only be discovered and requested after the main resource is downloaded and parsed. If redirection occurs, the process will be much longer. On smartphones, loading the main resource can contribute more than 50% of the browser delay. On average, getting the first data packet of the main resource takes 2 seconds under 3G network. If the main resource contains JavaScripts, the parsing of the main resource file can be further delayed, resulting in even longer time to discover subresources. Moreover, the dependencies between the resources will further serialize the resource loading operations [20].

In this work, we calculate the *browser delay* as follows: the starting point is when the user hits the "GO" button of the browser or clicks a URL to open a webpage. The end point is when the browser completely presents the requested webpage to the user, i.e., the browser's page loading progress bar indicates 100%. Such latency covers the time spent in all operations involved in opening a page, and can be unambiguously measured by keeping timestamps in the browser code. Though modern browsers utilize incremental rendering to display partially downloaded webpage to users, we do not consider partially displayed webpage as the metric because it is subjective how partial is enough to conclude the webpage is opened.

### 2.2 Related Work

Many have studied ways to improve browser speed, in particular resource loading. While only very few have specifically targeted mobile browsers, we discuss related work in terms of their approaches. Most proposals require infrastructure support, either from the web server or a proxy, e.g., thin-client approaches [3-6] and session-level techniques [7]. Web prefetching with infrastructure support is also widely studied, e.g. [8-11], and is shown to be effective in real world [21-24]. In a spirit similar to prefetching, *Crom* [25] speculatively runs JavaScript event handlers, prefetches the web data and pre-upload local files, also with server help. A recent protocol proposal, *SPDY* [12], improves the web performance by providing multiplexed streams, request prioritization, HTTP header compression, server push and server hint. It does so by adding a session layer atop of SSL and requires changes on both client and server. Though the approaches discussed above are effective, they are hard to deploy, are subject to the ability of the servers, cannot provide end-to-end security or has limited client JavaScript support.

Client-only solutions are attractive because they can be immediately deployed and work with existing web content. The authors of [16-18] sought to improve the client speed of compute-intensive operations in browser. As we showed in [2], their solutions will lead to negligible improvement in mobile browser speed. Existing client-solutions targeted at resource loading employ one or both of the following two approaches. *Browser caching* [26] is the most widely used client approach. As we will show in Section 4.1, caching is not effective for mobile browsers because of the long RTT and the large percentage of revalidations [27]. Web prefetching can also be implemented without server support. However, as we will show in Section 4.2, client-only prefetching introduces considerable unnecessary data usage with limited performance improvement because of the low prediction accuracy, which confirms previous observations on PCs [28].

### 2.3 LiveLab: Web Usage by 24 iPhone Users

Our work leverages web usage data collected from LiveLab [14], an unprecedented study of 24 iPhone 3GS users from February 2010 to February 2011. The 24 participants were recruited to have balanced gender, major, socioeconomic status to represent the

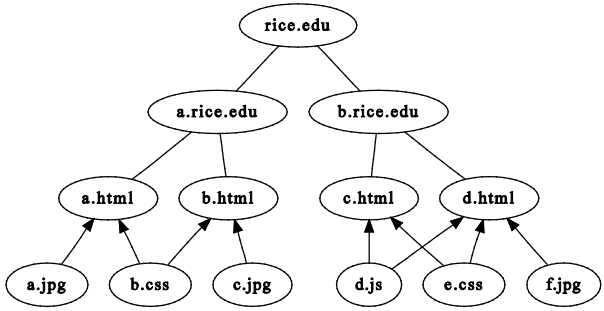

**Figure 2: Resource graph for simplified Rice University website. The arrows correspond to the dependency relationship between the webpage node and subresource node, i.e. the subresources can only be discovered after the main resource is parsed**

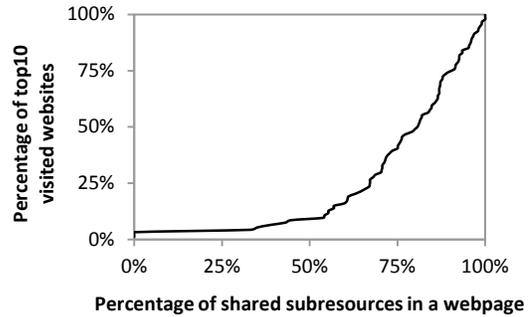

**Figure 3: CDF for average percentage of shared subresources in a webpage, i.e. subresources that are also needed by other webpages in the same website, for 94 web sites among the top 10 visited websites of each LiveLab user**

Rice University undergraduate population but not the general smartphone user population. All participants received unlimited data and were required to use the outfitted iPhones as his or her primary device. Almost all aspects of iPhone usage and context were collected by in-device, *in situ* programmable, logging software. The web usage data used in this work contains user id, timestamp and URLs of web pages visited. The top 10 visited web sites by each LiveLab user account for the majority (81%) of the user's webpage visits. Out of the top 10 of all 24 users, there are 94 websites, which will be used as benchmark web sites in this study. The LiveLab web usage data provide us a unique opportunity to understand mobile web browsing.

The 24 participants obviously cannot represent the general smartphone user population. However, they do provide us an important window into the latter. More importantly, most of our findings are not tied to the special demography of the 24 participants and we believe most, if not all, conclusions drawn in this paper regarding mobile browser performance should be applicable to a large fraction of the general population.

## 3. Mobile Web Browsing Characteristics

To study the effectiveness of client-only solutions for mobile browsers, we first study browsing behaviors of smartphone users and characterized websites visited using the LiveLab traces.

### 3.1 Characteristics of Websites

Since resource loading is the key to browser performance, it gains insight for improvement to examine how a web page needs many resources and how web pages from a web site may share resources. Toward this end, we represent each website, its subdomains, web pages, and subresources with a graph, called *resource graph*. Figure 2 shows an example of the resource graph for the simplified Rice University website. The resource graph has four types of nodes: *website node*, *subdomain node*, *webpage node* and *subresource node*. *Website node* is represented by the top two level domain names of the website. *Subdomain node* is a subdomain of the website. *Webpage* and *subresource nodes* are the real resources in the website and can be addressed by their URLs. The webpages mainly correspond to HTML files and the subresources mainly correspond to JavaScript, CSS, and image files.

The arrows between nodes in a resource graph denote the dependency relationship between the webpage node and subresource node. That is, the subresources can only be discovered after the main resource is parsed. Most of the dependencies occur between the webpage node and its subresource nodes. After executing some JavaScript and CSS files, the browser may discover and request new subresources. With a complete resource graph of a web site, we know which subresources are needed to open a webpage from the site.

While each website has its own complete resource graph, users can usually only see part of it, depending on which webpages the users visit. We download each LiveLab user's top 10 visited websites' homepages and the linked webpages, and then construct a partial resource graph for each website. Though the constructed resource graph is partial, we manually check that it represents the resource structure of the website. We have the following two observations.

First, *webpages from the same website often share a large portion of resources*. Those shared resources are the subresource nodes having multiple webpage nodes in the resource graph. Figure 3 shows the cumulative distribution function (CDF) for average percentage of shared subresources in a webpage, i.e. subresources that are also needed by other webpages in the same website, for top 10 visited websites. On average, 76% of the resources in one webpage are shared by at least one other webpage from the same website. This observation provides a key opportunity to improve the speed of opening a new webpage. After the user visits the website enough times and the resource graph is constructed, the browser can potentially predict the majority of the subresources needed for a new webpage visit, and thus speculatively load them (Section 5.2).

Second, *the structure of the resource graph can change over time*. New webpage subresource nodes can be added into the resource graph. A typical example is news website, which has changing content in the website all the time. Different website's resource graph changes in different frequency. For each LiveLab user's top 10 visited websites (in total 94 websites), 24 websites add new webpage nodes every a few hours or in even shorter period (fast changing); 13 websites add new webpage nodes daily; and 57 websites are stable and no new webpage nodes are added over a long period of time. Among the fast changing websites, 4% of the webpage nodes and 10% of the subresource nodes are replaced by new ones every hour. Among the unchanged webpage nodes in fast changing websites, 26% of them have new subresource nodes, in which 11% of those subresource nodes are replaced with new ones. This observation raises challenges to solutions that leverage the resource graph, because temporal change of website's resource graph is hard to be captured by the client timely. However, our speculative mobile browser design, Tempo, can deal with the temporal change well and reduce the browser delay by 1 second (Section 6).

## 3.2 User Browsing Behavior on Smartphone

Understanding the browsing behavior of smartphone users helps us to study the effectiveness of client-only solutions and better design Tempo. We have four interesting findings. First, *for a given smartphone user, the total number of frequently visited websites is usually small*. The user's top 10 visited websites account for 81% of his/her total webpage visits. Therefore, it is reasonable to focus on the resource loading optimization for the webpages that belong to the top 10 visited websites.

Second, *across different users, the web usage is diverse*. Approximately three (both average and median) of the users' top-10 websites were shared by the all-users-combined top-10 list. Therefore, resource loading optimization should target at different sets of websites for different users, which can be easily achieved by client-only solutions.

Third, *the majority of the webpage visits are new visits*. On average, 75% of the webpages visited are new visits. The high new webpage visit rate is one of the reasons that client-only web prefetching has poor performance on mobile browsers (Section 4.2).

Fourth, *though users tend to visit new web pages, the browser is likely to request a similar set of subresources*. On average, only 35% of the subresources requested are new subresources. The reason is that webpages in the same website share subresources, as discussed in the previous section. Therefore, subresources can be much more predictable than webpages. This is the key reason that Tempo outperforms client-only web prefetching.

## 4. Effectiveness of Client Only Approaches

Driven by findings presented above, we next examine three orthogonal client-only approaches that speed up resource loading. With *caching*, browser saves the subresources of previously visited webpages locally and reduces the resource loading time if the same subresources are requested again. *Web Prefetching* predicts which webpage a user is likely to visit and downloads its resources beforehand; it minimizes the resource loading time if the user does visit a prefetched page. We show both caching and prefetching are limited for mobile browsers, and show how a new, orthogonal approach, called *speculative loading*, can be much more effective. We reported the early results of our study on browser caching in a workshop paper [27].

### 4.1 Caching

Caching is a well-known approach to fight I/O bottlenecks. A browser stores frequently used web resources locally to save the RTT and bandwidth. But resources with "no-store" in the cache-control header field cannot be stored in the browser cache.

A cached resource can have two states: fresh or expired. The browser can return a fresh resource in response to the request without contacting the server. The browser needs to revalidate an expired resource with the origin server to see if the resource is still usable. If it is usable, the server will not send back the entire resource file. Resources with "no-cache" in the cache-control header field can be actually cached but they immediately expire. Both HTTP and HTTPS resources can be cached but their expiration time is indicated in the header by the server.

A working browser cache is a mixture of fresh and expired resources. Because a large portion of mobile web resources either cannot be cached or have a short expiration time, caching brings little benefit to mobile browsing. Usually, by revalidating expired resources with the server, the browser avoids re-fetching resources if local copies are still usable. However, revalidation cannot hide the extra network RTT, and RTT is the most important factor to mobile browser delay [2]. As a result, latencies in revalidations make caching ineffective for mobile browsers.

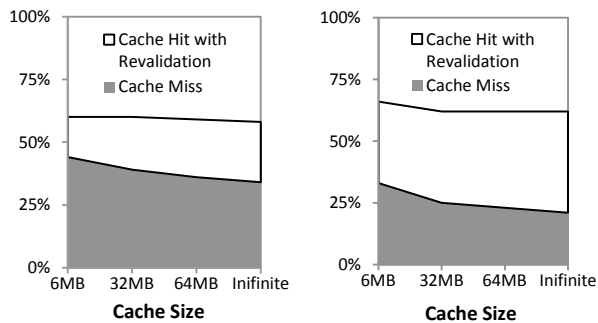

**Figure 4: Cache simulation results for the webpages from all websites (Left) and top 10 visited websites (Right)**

We experimentally show how excessive revalidations outweigh the benefit of caching, with the LiveLab traces[14]. Firstly, we download all the resources of the webpages with header information. We simulate the cache behavior of the mobile browser by replaying each user's browsing history using the file and HTTP header information of each resource. We repeat the simulation with four cache sizes: 6MB, 32MB, 64MB and infinite. Note that Android Gingerbread browser's default cache size is 6MB. We have to exclude about 32% of the webpage visits, including visits to pages that no longer exist (39%) and to HTTPS webpages (61%) that require user login. Excluding HTTPS webpages does not bias the results much because most of their resources' expiration time is not different from their HTTP counterpart.

As shown in Figure 4, our simulation results show that 60% of resource requests incur network activities with a 6 MB cache. Network activity is required when a requested resource is not in the cache, or cache miss. It may also be required even if the resource is in the cache: when a cached resource is expired but requested, the browser still has to contact the server to revalidate it. As is apparent from Figure 4, the effectiveness of caching is even lower for the top 10 web sites of each user: 70% of resource requests incur network activities, in which half of them are due to revalidations.

Note that increasing the cache size will not help much. Figure 4 shows that the small size (6MB) of current browser cache incurs only 10% more cache misses than infinite cache. And 58% of resource requests incur network activities with an infinite size cache, which is close to the percentage with 6MB cache size (60%). Therefore, a larger cache will not bring much benefit.

In summary, our results show that the benefit of caching is marginal because it can do little in loading resources whose cached copies expire quickly: revalidation saves the bandwidth usage in this case, but cannot hide the network RTT, which is the most important factor to mobile browsers' performance [2].

### 4.2 Web Prefetching

We believe that client-only web prefetching [9-11] is harmful to mobile web browsing, because it results in significant additional data usage with very little improvement. Web prefetching predicts the webpages that will be visited by the user and downloads their resources beforehand. When the user actually visits a predicted webpage, its resources are already available locally. Most solutions of web prefetching are intended for PC browsers and involve infrastructure support to aggregate behaviors of many users. We showed that on smartphones, client-only web prefetching is ineffective because web prefetching cannot predict URLs that have never been visited before; and on average, 75% of the web pages visited are new visits, as shown in Section 3.2.

Table 1: Upper bound of the browser delay reduction from speculative loading under different cache states (in ms)

| Sites | Fresh Cache | | | | Expired Cache | | | | Empty Cache | | | |
|---|---|---|---|---|---|---|---|---|---|---|---|---|
| | Legacy | Speculate | Reduction | | Legacy | Speculate | Reduction | | Legacy | Speculate | Reduction | |
| ESPN | 4557 | 4557 | 0 | 0% | 6702 | 4622 | 2080 | 31% | 7143 | 4622 | 2521 | 35% |
| CNN | 2382 | 2382 | 0 | 0% | 4869 | 2884 | 1985 | 41% | 6300 | 4315 | 1985 | 32% |
| Google | 2162 | 2131 | 31 | 1% | 3363 | 2131 | 1232 | 37% | 3661 | 2223 | 1438 | 39% |
| Yahoo Mail | 3199 | 3199 | 0 | 0% | 4333 | 3199 | 1134 | 26% | 4341 | 3199 | 1142 | 26% |
| Weather | 3645 | 3608 | 37 | 1% | 6294 | 3608 | 2686 | 43% | 6349 | 3608 | 2741 | 43% |
| Craigslist | 1926 | 1920 | 6 | 0% | 3034 | 1920 | 1114 | 37% | 3103 | 1920 | 1183 | 38% |
| Neopets Games | 3605 | 3605 | 0 | 0% | 11505 | 9002 | 2503 | 22% | 11843 | 9340 | 2503 | 21% |
| Varsity Tutors | 3313 | 3313 | 0 | 0% | 8410 | 6596 | 1814 | 22% | 9219 | 7405 | 1814 | 20% |
| Ride METRO | 3826 | 3826 | 0 | 0% | 8266 | 5560 | 2706 | 33% | 8774 | 6068 | 2706 | 31% |
| Rice Registrar | 3351 | 3351 | 0 | 0% | 5865 | 3541 | 2324 | 40% | 6427 | 3541 | 2886 | 45% |
| **Average** | **3197** | **3189** | **7** | **0%** | **6264** | **4306** | **1958** | **33%** | **6716** | **4624** | **2092** | **33%** |

To quantitatively demonstrate this ineffectiveness, we evaluate client-only web prefetching using the LiveLab traces. We simulate the web prefetching algorithm presented in [11], called *mostpopular*. It uses the popularity ranking of user's past requests to predict future requests. We also borrow the metrics, *hit ratio* and *usefulness*, from [11]. The *hit ratio* is defined as the number of webpages that are predicted and also actually requested to the number of predicted webpages. It represents the accuracy of the prediction. High hit ratio means low unnecessary data usage. The *usefulness* is defined as the number of webpages that are predicted and also actually requested to the number of actually requested webpages. It represents the coverage of the prediction. High usefulness means high average speedup.

With one month training period, the hit ratio is 16% and the usefulness is 1% on average among 24 iPhone users. Such low hit ratio and usefulness lead to considerable unnecessary data usage yet very limited speed improvement. With a very generous assumption that the prefetched content is cached and will not expire before actual visit, the upper bound of the browser delay reduction from the most-popular web prefetching algorithm is 1%. And the unnecessary data usage accounts for 84% of the total prefetched data.

One may think that prefetched subresources for one webpage may help make other webpages from the same web site faster because subresources are shared by webpages from the same web site as shown in Section 3.1. Unfortunately, this is usually not the case because many resources are either not in the cache or their cached copies expire quickly, as shown in Section 4.1. In contrast, speculative loading solves this problem by loading the resources only *after* the user requests a webpage's URL.

## 4.3 Speculative Loading

Seeing the failures of caching and prefetching, we propose a third, orthogonal approach called *speculative loading* that loads subresources for a web page along with the main resource file *after* a user provides the web URL.

Essentially, speculative loading predicts which subresources to load based on a resource graph of the website constructed using knowledge of the website collected from the past. It leverages the many concurrent connections (e.g. 4 for Android Gingerbread) available to modern browsers to concurrently load subresources along with the main resource. Unlike caching, speculative loading will revalidate expired resources and load evicted resources concurrently with loading the main resource, thus keeping most subresources fresh in cache when the browser actually requests them. Unlike web prefetching, speculative loading predicts which resources a web page may need, instead of which webpage the user may visit.

### 4.3.1 Upper Bound of Improvement

The key to the effectiveness of speculative loading is *subresource prediction*. By assuming 100% hit ratio and 100% usefulness for subresource prediction, we are able to derive the upper bound of the browser delay reduction from speculative loading. We will show in Section 6 that the performance of speculative loading is close to this upper bound in practice. Here we examine the browser delays for the homepages of top visited websites from LiveLab traces under three different cache states: fresh, expire, and empty. With fresh cache, if a requested resource is cached, the browser will use the cached copy without any network activity. With expired cache, if a requested cache is cached, the browser still needs to revalidate it with the server. With empty cache, the browser needs to load every resource file from the server.

Table 1 shows the upper bound of the browser delay reduction. We measure the browser delays of legacy loading with empty cache on Samsung Galaxy S II in 3G network provided by U.S. wireless carrier AT&T. Then we simulate the browser delays in other columns by applying what-if analysis as described in [2]. In summary, what-if analysis tries to derive the overall performance gain if a browser operation is accelerated. To accurately predict the impact of accelerating all instances of any operation, we scale the execution time of each instance of such an operation. All operation instances depending on it will thus be executed earlier, resulting shorter browser delay.

We have three observations. (*i*) The average browser delay reductions are 33% (~2 seconds) for expired and empty cache. The reduction comes from the time waiting for the main resource to discover the subresources. (*ii*) There is nearly no reduction for fresh cache because all the subresources are available locally already. There is no advantage of discovering and loading subresources speculatively. (*iii*) The upper bound of browser delay reduction by speculative loading for a realistic cache can be estimated to be around 1.4 (22%) seconds because, when a webpage from top10 visited websites is visited, 70% of its subresources needed by a webpage are either expired or not in the cache, as shown in Section 4.1.

*4.3.2 Predicting Server vs. Predicting User*

Speculative loading shows more promise than web prefetching. The upper bound of the browser delay reduction from speculative loading (22%) is one order of magnitude larger than the upper bound of reduction from web prefetching (1%). Moreover, by applying the design discussed in Section 5, speculative loading will consume much lower unnecessary data usage with 65% hit ratio as will be evaluated in Section 6, comparing to 16% hit ratio for web prefetching.

There is a fundamental reason that speculative loading can be much more effective than web prefetching: predicting server behavior is much easier than predicting user behavior. Speculative loading predicts the subresources needed by a webpage, which is server behavior prediction. Web prefetching predicts the next visited webpage by the user, which is user behavior prediction. Server behavior prediction can achieve high accuracy because webpages in the same website share subresources, as discussed in Section 3.1. On the contrary, user behavior prediction is limited because 75% of the visited webpages are new visits, as presented in Section 3.2. To predict server structure, the browser needs to map the resource graph of each website on the smartphone and we will discuss the detailed design in Section 5.

## 4.4 Upper Bound for Client-Only Solutions

Existing two client-only approaches are limited because of two reasons, respectively. First, a large portion of mobile web resources are either not in the cache or their cached copies quickly expire, which makes caching ineffective. Second, mobile browsing behaviors are not very predictable, which makes client-only web prefetching harmful. Our proposed approach, speculative loading, addresses those two limitations by speculatively revalidating expired resources and loading evicted resources, and by predicting server behavior instead of predicting user behavior.

Speculative loading has reached the upper bound of improvement for client-only solutions, i.e. 1.4 seconds as shown by us empirically. The reason is that the request-response model of HTTP protocol [15] requires at least one request for each resource needed and the loading procedure is already fully parallel with speculative loading. In practice, it is difficult to completely predict what resources are needed for a webpage based on its user's history. Our speculative mobile browser design, Tempo, can reduce the browser delay by 1 second, as will be evaluated in Section 6.2, a result close to the upper bound.

According to our previous work [2], better hardware can also speed up resource loading by providing faster OS services and network stack. The browser speedup from hardware improvement is orthogonal to the upper bound of improvement achieved by the client-only approaches discussed above.

## 5. Tempo: A Speculative Mobile Browser

We now describe *Tempo*, our mobile browser design that seeks to realize the potential of speculative loading. As illustrated in Figure 5, Tempo is realized by adding a module under the middle layer in Android Gingerbread browser. The middle layer connects the browser engine (WebKit [29]) and the network service provided by the smartphone. It also handles the communication between the browser user interface and WebKit, and manages caches, cookies and plug-ins.

Tempo has four components. *Metadata repository* stores each website's resource graph and makes speculative loading possible. S*peculative loader* predicts the subresources needed based on the information provided by metadata repository and loads the pre-dicted subresources speculatively for every webpage visit. *Update service* updates *metadata repository* with the new resource information after the webpage is open and trims the stale nodes in *metadata repository*. The last component is *temporary cache*, which stores the resources that cannot be stored in the cache temporarily (those with "no-store" in cache-control header). We will discuss the details of each component as follows.

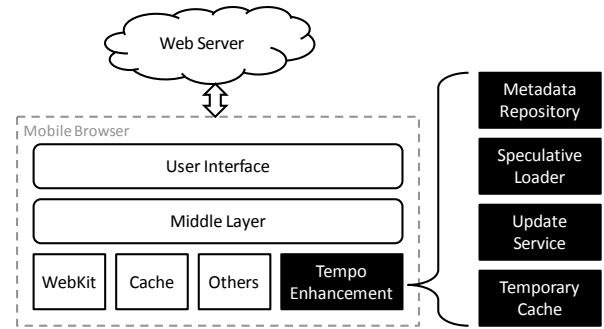

**Figure 5: Tempo, a speculative mobile browser. Black components are new additions**

### 5.1 Metadata Repository

Metadata repository is a key-value store, as shown in Figure 6. The key is the website and the value is the website's resource graph, which is discussed in Section 3.1. Each node in the resource graph has several fields, e.g. type, URL, last visit time, children, parents, and number of visits. The actual content is not stored in the resource graph.

Metadata repository has two advantages. Firstly, it relates the resources in each website in the corresponding resource graph. When visits occur, the browser knows which subresources are needed even before downloading and parsing the main resource file. This makes speculative loading possible. In contrast, caching provides no relation information among the cached resources. Secondly, metadata repository takes little storage on the smartphone (only several hundred KB) because each node in the resource graph is represented by the URL instead of the actual content.

Metadata repository is stored in the flash storage. Loading or saving it will not affect the browser delay because it will be loaded into the memory when the browser starts and will be saved to the flash storage after each webpage is open.

### 5.2 Speculative Loader

Speculative loader takes the webpage's URL as the input right after the user enters or clicks the URL of that webpage, finds the corresponding resource graph from the metadata repository, predicts the subresources needed for that webpage based on the resource graph, and loads those subresources speculatively if they are not in the cache or expired. It not only handles webpage revisits but also handles new webpage visits. Note that speculative loading for new webpage visits is very important and cannot be ignored for mobile browsers, because a large portion of webpage visits are new visits as discussed in Section 3.1. In contrast, web prefetching relies on past history and cannot benefit new visits. That's one of the reasons that web prefetching has poor performance on smartphones.

The detailed subresource prediction algorithm is illustrated in Figure 7. If the webpage visit is a revisit, speculative loader can find the corresponding webpage node in the resource graph and thus all its child subresource nodes are the subresources the webpage needs. If the webpage visit is a new visit, no corresponding webpage node is stored in the resource graph yet. Speculative loader predicts the subresources' URLs according to the shared subresource nodes because subresources are heavily shared across multiple pages of the same website, as discussed in Section 3.1.

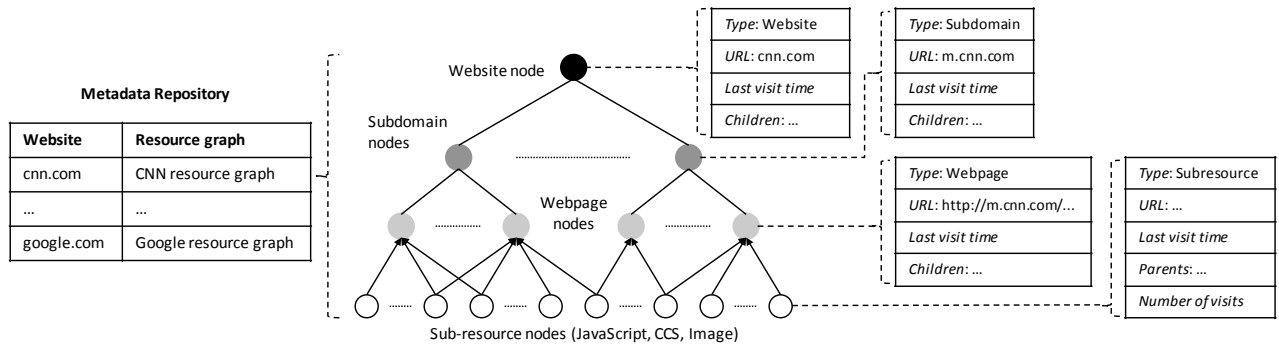

**Figure 6: Metadata repository, a key-value store where keys are websites and values are websites' resource graphs**

To maximize the prediction accuracy and coverage, speculative loader judiciously prioritizes the candidate subresources by sorting them according to their number of parents (large to small), file types (JS to CSS to image), number of visits (large to small) and URL length (short to long), as indicated by the function Sort() in Figure 7. If it is webpage new visit, speculative loader only chooses the ones with high priority as the predicted subresources, i.e. the subresource nodes that are shared by more webpage nodes). JavaScript and CSS files have higher priority than images because they may further request subresources and scripting may block later executions. Long URLs has higher chance to contain session dependent string, which makes the URL useless next time. So long URLs have low priority.

To reduce the unnecessary data usage, speculative loader loads the predicted subresources adaptively. When the number of predicted resources is more than the number of allowed concurrent connections, the resources with higher priority will be requested immediately and other resources will be put into a waiting queue. If main resource file is downloaded and parsed before the waiting resources are actually requested, the waiting queue will be updated with the actually needed resources, reducing unnecessary data usage from prediction misses.

### 5.3 Update Service

Update service constructs and modifies the resource graphs in the metadata repository. There are two major operations performed on the nodes in the resource graph: *update* and *trim*. *Update* operation adds a node if the node does not exist in the resource graph or updates the information stored in the node if the node exists in the resource graph already. *Trim* operation removes the nodes that are not visited for more than one month from the resource graph.

After a webpage is open, update service *updates* the webpage nodes, its subresource nodes, the corresponding subdomain node and website node in the resource graph. Some webpages dynamically request subresources after a webpage is open, e.g. by using AJAX. Update service can also capture those requests and update the subresource nodes accordingly. Every day, update service *trims* resource graph and remove the stale nodes, whose last visit time is older than a month. *Trimming* resource graph keeps the user viewed website resource structure up-to-date and limits the storage metadata repository takes.

```
Input: webpage URL
Output: predicted subresources' URLs
SubresourcePrediction(url):
  candidates = []             // subresources
  webpage_node = get_webpage_node(url)
  if webpage_node != NULL:    // webpage revisit
    candidates = children_of_webpage_node
    sorted_candidates = Sort(candidates)
    return sorted_candidates
  else:                       // webpage new visit
    subdomain_node = get_subdomain_node(url)
    if subdomain_node != NULL:
      candidates = subres_nodes_of_the_subdomain
    else:
      candidates = subres_nodes_of_the_website
    sorted_candidates = Sort(candidates)
    num_predicted = avg_num_of_webpage_children
    return sorted_candidates[0:num_predicted]
```

**Figure 7: Pseudo Code of Subresource Prediction**

### 5.4 Temporary Cache

The purpose of the temporary cache is to store the resources that have "no-store" in their cache-control header field temporarily. Those files should not be stored in the cache. When speculative loader loads predicted resources, resources with "no-store" in their cache-control header field will be saved to temporary cache and other resources will be saved to normal cache. Later when WebKit actually requests the speculatively loaded resources, it will get them either from normal cache or temporary cache. After the webpage is open, all the resources in temporary cache will be deleted.

## 6. Evaluation

We evaluate the Tempo design through trace-based simulation, lab experiments and a field trial. The evaluation shows that the subresource prediction has high accuracy and coverage, resulting in 1 second (~20%) of browser delay reduction with small overhead.

### 6.1 Subresource Prediction Performance

We firstly evaluate how good the subresource prediction is and how long Tempo needs to learn to make good predictions, based on the LiveLab traces. There are two metrics: *hit ratio* and *useful-* *ness*. As mentioned in Section 4.2, *hit ratio* represents the accuracy of the prediction and *usefulness* represents the coverage of the prediction.

Figure 8 shows the weekly and monthly hit ratio and usefulness of subresource prediction, respectively. We can see that the first week's hit ratio (50%) and usefulness (56%) are already much higher than web prefetching shown in Section 4.2. The highest hit ratio (65%) and usefulness (73%) are reached in the $3^{rd}$ month. Interestingly, they drop slightly at week 4, 5, 12 and month 4, 5, largely when the LiveLab users visited a different set of websites around holidays and school breaks and Tempo takes time to construct the resource graph for the new websites.

## 6.2 Lab Experiments

We now evaluate how the subresource prediction performance is translated into browser delay reduction through lab-based experiments. For the experiments, we port Tempo to Samsung Galaxy S II smartphone that runs Android Gingerbread [30]. The code of Tempo is instrumented to record webpage delay efficiently. All experiments use 3G network provided by AT&T, a major U.S. carrier, at our lab on Rice campus where 3G signal strength is strong.

### 6.2.1 Revisits

Firstly, we show that Tempo can reduce browser delays of webpage revisits to very close to the upper bound presented in Section 4.3. We use the homepages of the websites from Table 1 in the experiment. We firstly open the URLs in the browser once to warm up the cache and construct the resource graph. Then we open the URLs one by one for five times and calculate the average browser delay. Even though all the webpage visits are revisits in the experiment and we have minimized the time interval between revisits, there can still be cache misses due to the dynamic and/or session dependent content in the web. Subresource prediction cannot predict all the subresources needed, either. On average, the hit ratio is 65% and the usefulness is 72%, which are close to the prediction accuracy and coverage evaluated in the previous section.

We compare the browser delays between legacy loading and Tempo with three different cache states, similar to what used in Section 4.3, i.e., fresh, expired, and empty. The browser is modified to always revalidate the resources stored in the cache under expired cache and clears the cache before each webpage visit under empty cache.

Table 2 shows the browser delays of webpage revisits under different cache states with the WebKit browser without speculative loading (*Legacy*) and Tempo. With fresh cache, the browser delays of Legacy and Tempo are close because most of the subresources are available locally. With expired cache, Tempo reduces 25% (1445ms) of browser delay on average. Tempo also reduces 24% (1464ms) of browser delay under empty cache. Since 70% of the requested resources of a webpage from top 10 visited websites are either expired or not in the cache, as mentioned in Section 4.1, we estimate tempo can reduce the browser delay by around 1 second or ~20% with a realistic cache. This one second browser delay reduction is also confirmed by our field trial, which will be discussed in the next section.

The browser delay reduction for each website mainly comes from the time waiting for the main resource to discover the subresources. Thus the content richness of the webpage (number of subresources) does not have a direct influence on browser delay reduction and most of the reductions in Table 2 are close to each other (~1.4 second). The time spent to download and parse the main resource affects the discovery time of subresources and there are two main factors: *(i)* main resource redirection delays the download of main resource, e.g. Weather website; *(ii)* JavaScript execution can delay main resource parsing, e.g. Varsity Tutors website. Since Tempo eliminates the resource dependencies, it can provide more browser delay reduction for websites that have previous two limiting factors.

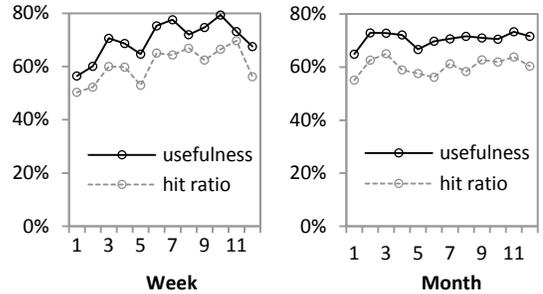

**Figure 8: Hit ratio and usefulness of subresource prediction for the first 12 weeks (Left) and the entire year (Right). Each data point is the average value across 24 LiveLab users**

The browser delay reduction of Tempo is very close to the upper bound presented in Section 4.3. Under expired or empty cache, Tempo can reduce around 1.4 second, which is 70% of the upper bound (reduce 2 seconds) we can get. For realistic cache, Tempo can reduce around 1 second, which is 71% of the upper bound with realistic cache (around 1.4 second). By achieving its design goal, Tempo essentially keeps most subresources fresh in cache when the browser requests them. Table 2 shows that the average browser delay of Tempo under expired and empty cache (4597ms and 5222ms) is only 3% and 17% larger than that of Legacy under a fresh cache (4446ms), which is the ideal case.

### 6.2.2 New Visits

Tempo can also greatly reduce the browser delay for new webpage visits, which account for 75% of the total web page visits in LiveLab traces. In the experiment, we use the websites in Table 1. We firstly open the homepage of the website in the browser once to warm up the cache and construct the resource graph. Then we navigate to five other webpages in the same website and then calculate the average browser delay. The browser delay for the homepage is not counted. Even though new webpages are used, and thus the subresources needed by the webpage will be different, subresource prediction can still predict some of the subresources needed, from the shared subresource nodes. On average, the hit ratio is 50% and the usefulness is 67%, which is only slightly lower than the prediction accuracy and coverage evaluated in the previous section.

With similar cache states, we compare the browser delays between Legacy and Tempo. Table 3 shows the browser delays of new visits to webpages of different websites. Under fresh cache, Legacy and Tempo exhibit similar browser delay. Under expired and empty cache, on average, Tempo incurs 20% (960ms) and 17% (1119ms) less browser delay than the Legacy, respectively. Notice that the browser delay of Tempo under expired cache (3616ms) is even 19% smaller than that of Legacy under fresh cache (4465ms). The reason is that Tempo can effectively revali-date the expired subresources, warm up the TCP connection and thus download new subresources much faster.

**Table 2: Browser delay reduction from speculative loading for *webpage revisits* under different cache states (in ms)**

| Sites | Fresh Cache | | | | Expired Cache | | | | Empty Cache | | | |
|---|---|---|---|---|---|---|---|---|---|---|---|---|
| | Legacy | Tempo | Reduction | | Legacy | Tempo | Reduction | | Legacy | Tempo | Reduction | |
| ESPN | 3491 | 3602 | -111 | -3% | 6748 | 5372 | 1376 | 20% | 7031 | 5322 | 1709 | 24% |
| CNN | 4873 | 4507 | 366 | 8% | 5992 | 4274 | 1718 | 29% | 6346 | 5307 | 1039 | 16% |
| Google | 2407 | 2842 | -435 | -18% | 3411 | 3073 | 338 | 10% | 3932 | 3257 | 675 | 17% |
| Yahoo Mail | 3239 | 3472 | -233 | -7% | 5083 | 3265 | 1818 | 36% | 5083 | 3442 | 1641 | 32% |
| Weather | 5055 | 4559 | 496 | 10% | 6109 | 3835 | 2274 | 37% | 7167 | 4716 | 2451 | 34% |
| Craigslist | 3123 | 2400 | 723 | 23% | 3648 | 2089 | 1559 | 43% | 3677 | 2470 | 1207 | 33% |
| Neopets Games | 9041 | 9076 | -35 | 0% | 10639 | 9280 | 1359 | 13% | 10660 | 10220 | 440 | 4% |
| Varsity Tutors | 5969 | 5384 | 585 | 10% | 8516 | 6677 | 1839 | 22% | 9987 | 7914 | 2073 | 21% |
| Ride METRO | 4220 | 3801 | 419 | 10% | 6109 | 4620 | 1489 | 24% | 6945 | 5488 | 1457 | 21% |
| Rice Registrar | 3046 | 3609 | -563 | -18% | 4169 | 3489 | 680 | 16% | 6027 | 4084 | 1943 | 32% |
| **Average** | **4446** | **4325** | **121** | **1%** | **6042** | **4597** | **1445** | **25%** | **6686** | **5222** | **1464** | **24%** |

**Table 3: Browser delay reduction from speculative loading for *new webpage visits* under different cache states (in ms)**

| Sites | Fresh Cache | | | | Expired Cache | | | | Empty Cache | | | |
|---|---|---|---|---|---|---|---|---|---|---|---|---|
| | Legacy | Tempo | Reduction | | Legacy | Tempo | Reduction | | Legacy | Tempo | Reduction | |
| ESPN | 3152 | 2587 | 565 | 18% | 3163 | 2788 | 375 | 12% | 6162 | 4205 | 1957 | 32% |
| CNN | 2994 | 3328 | -334 | -11% | 3519 | 2438 | 1081 | 31% | 7091 | 6054 | 1037 | 15% |
| Google | 2982 | 2295 | 687 | 23% | 2376 | 2492 | -116 | -5% | 4638 | 2945 | 1693 | 37% |
| Yahoo Mail | 5222 | 5282 | -60 | -1% | 4472 | 3162 | 1310 | 29% | 5572 | 5047 | 525 | 9% |
| Weather | 5180 | 3763 | 1417 | 27% | 3757 | 2682 | 1075 | 29% | 5357 | 5244 | 113 | 2% |
| Craigslist | 1203 | 1210 | -7 | -1% | 2624 | 1848 | 776 | 30% | 5163 | 3463 | 1700 | 33% |
| Neopets | 10105 | 9795 | 310 | 3% | 7326 | 7038 | 288 | 4% | 6914 | 6623 | 291 | 4% |
| Varsity Tutors | 7126 | 8013 | -887 | -12% | 10598 | 7437 | 3161 | 30% | 14921 | 12674 | 2247 | 15% |
| Ride METRO | 2759 | 3460 | -701 | -25% | 3352 | 2602 | 750 | 22% | 6829 | 6171 | 658 | 10% |
| Rice Registrar | 3929 | 3708 | 221 | 6% | 4570 | 3672 | 898 | 20% | 6506 | 5534 | 972 | 15% |
| **Average** | **4465** | **4344** | **121** | **3%** | **4576** | **3616** | **960** | **20%** | **6915** | **5796** | **1119** | **17%** |

## 6.3 Field Trial

We also conducted a field trial to study the performance of Tempo browser. In the field trail, two Samsung Galaxy S II smartphones are used by two participants. Both smartphones are running Android Gingerbread [30] with Tempo browser and using the 3G network provided by U.S. wireless carrier AT&T. The field trial lasted for two weeks for each participant. Speculative loading was enabled in the first week, but disabled in the second week. The cache was cleared before the field trial and was never cleared during the field trial.

The results are shown in Figure 9. The average browser delay of the $2^{nd}$ week is 433ms longer than the delay of the $1^{st}$ week and 1424ms longer than the delay of the last two days in the $1^{st}$ week (day 6 & 7). The results show that once Tempo has warmed up the cache and constructed the resource graph, it outperforms Legacy by over one second, which is consistent with our findings from the lab experiments described early. The results also indicate that Tempo is effective with just several days' training.

## 6.4 Overhead

Tempo incurs very small overhead of three types: performance, data usage, and storage usage. Tempo incurs performance overhead when a predicted subresource is not actually needed. The predicted subresource will occupy a TCP connection, making real needed subresources wait for available connections. We have minimized this overhead by prioritizing the predicted subresources and loading them adaptively, as discussed in Section 5.2. From the experiments presented in previous section, we can also see that the benefit from Tempo outperforms the overhead.

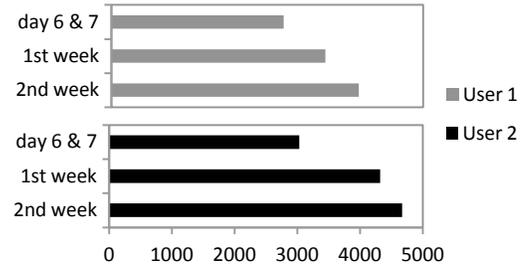

**Figure 9: Average browser delays (ms) of the field trial in different periods. Speculative loading is enabled in the 1st week and disabled in the 2nd week**

Tempo also incurs data usage overhead when a predicted subresource is not actually needed. Higher hit ratio will incur less additional data usage, as discussed in Section 4.2. Resource prediction in Tempo achieves hit ratio as high as 65%, as presented in Section 6.1, which is much higher than web prefetching (16%). Though 35% of the predicted subresources are not actually needed by current webpage, the actual unnecessary data usage are usually even lower because of three reasons: (*i*) the predicted subresources are loaded adaptively, which minimizes the data usage overhead; (*ii*) the predicted subresources are widely shared by different webpages in the website, so the overhead is amortized by other webpages; (*iii*) the predicted subresources are visited before and they may be still in the cache, resulting in little network traffic for expired resource or even no network traffic for fresh resource. We counted the data usage overhead in the field trial, to be only 0.7 MB per week.

Tempo incurs storage usage overhead by constructing and storing metadata repository on the smartphone, which requires additional disk usage. However, Tempo takes little additional storage on the smartphone because metadata repository does not store actual content, as discussed in Section 5.1. For the 24 LiveLab users, one year's metadata repository takes only 165KB on average (max 576KB), which is negligible in view of what is available to modern smartphones.

## 7. Conclusions

Of solutions for browser speed improvement, client-only ones are immediately deployable, scalable, and secure. It has been well known that client-only solutions are not as effective in improving speed as ones with infrastructure support. Leveraging an unprecedented mobile web usage data set, our work provides the first comprehensive treatment regarding the effectiveness of client-only solutions.

We demonstrate the ineffectiveness of browser caching and client-only web prefetching on mobile browsers. Caching is not effective because of the large portion of the resources that are either not in the cache or their cached copies quickly expire. Client-only web prefetching is harmful because it results in huge additional data usage with limited improvement.

In order to address the limitations of previous two approaches, we propose speculative loading, a client-only approach that predicts the subresources of its webpages when the user provides the webpage's URL and then speculatively loads the predicted subresources. Our implementation of speculative loading, Tempo, can reduce browser delay by 1 second (~20%) under 3G network.

Finally, we empirically show that the upper bound of browser delay reduction for client-only solutions is 1.4 second. Our result suggests that it is imperative to involve the infrastructure in further improving mobile browser performance.

## 8. REFERENCES

[1] R. Kohavi, R. M. Henne, and D. Sommerfield, "Practical guide to controlled experiments on the web: listen to your customers not to the hippo," in *Proceedings of the 13th ACM SIGKDD international conference on Knowledge discovery and data mining*, 2007.
[2] Z. Wang, X. Lin, L. Zhong, and M. Chishtie, "Why are web browsers slow on smartphones?," in *Proceedings ACM Int. Workshop on Mobile Computing Systems and Applications (HotMobile)*, 2011.
[3] Opera Mini: http://www.operamini.com/.
[4] Skyfire: http://www.skyfire.com/.
[5] A. M. Lai, J. Nieh, B. Bohra, V. Nandikonda, A. P. Surana, and S. Varshneya, "Improving web browsing performance on wireless pdas using thin-client computing," in *Proceedings of the 13th international conference on World Wide Web*, 2004.
[6] J. Kim, R. A. Baratto, and J. Nieh, "pTHINC: a thin-client architecture for mobile wireless web," in *Proceedings of the 15th international conference on World Wide Web*, 2006.
[7] P. Rodriguez, S. Mukherjee, and S. Ramgarajan, "Session level techniques for improving web browsing performance on wireless links," in *Proceedings of the 13th international conference on World Wide Web*, 2004.
[8] Amazon Silk: http://amazonsilk.wordpress.com/.
[9] V. N. Padmanabhan and J. C. Mogul, "Using predictive prefetching to improve World Wide Web latency," *SIGCOMM Comput. Commun. Rev.,* vol. 26, pp. 22-36, 1996.
[10] L. Fan, P. Cao, W. Lin, and Q. Jacobson, "Web prefetching between low-bandwidth clients and proxies: potential and performance," in *Proceedings of the ACM SIGMETRICS international conference on Measurement and modeling of computer systems*, 1999.
[11] C. Bouras, A. Konidaris, and D. Kostoulas, "Predictive Prefetching on the Web and Its Potential Impact in the Wide Area," *World Wide Web,* vol. 7, pp. 143-179, 2004.
[12] The Chromium Projects, "SPDY: An experimental protocol for a faster web," http://www.chromium.org/spdy.
[13] NYTimes, "Amazon Cloud Failure Takes Down Web Sites," http://bits.blogs.nytimes.com/2011/04/21/amazon-cloud-failure-takes-down-web-sites/.
[14] C. Shepard, A. Rahmati, C. Tossell, L. Zhong, and P. Kortum, "LiveLab: Measuring Wireless Networks and Smartphone Users in the Field," in *Proc. Workshop on Hot Topics in Measurement & Modeling of Computer Systems*, June 2010.
[15] "Hypertext Transfer Protocol -- HTTP/1.1," http://www.ietf.org/rfc/rfc2616.txt.
[16] C. Stockwell, "IE8 Performance," http://blogs.msdn.com/b/ie/archive/2008/08/26/ie8-performance.aspx, 2008.
[17] L. A. Meyerovich and R. Bodik, "Fast and parallel webpage layout," in *Proc. Int. Conf. World Wide Web (WWW)* Raleigh, North Carolina, USA: ACM, 2010.
[18] K. Zhang, L. Wang, A. Pan, and B. B. Zhu, "Smart caching for web browsers," in *Proc. Int. Conf. World Wide Web (WWW)* Raleigh, North Carolina, USA: ACM, 2010.
[19] J. Huang, Q. Xu, B. Tiwana, Z. M. Mao, M. Zhang, and P. Bahl, "Anatomizing application performance differences on smartphones," in *Proc. ACM/USENIX Int. Conf. Mobile Systems, Applications, and Services (MobiSys)* San Francisco, California, USA: ACM, 2010.
[20] Z. Li, M. Zhang, Z. Zhu, Y. Chen, A. Greenberg, and Y.-M. Wang, "WebProphet: automating performance prediction for web services," in *Proceedings of the 7th USENIX conference on Networked systems design and implementation*.
[21] Mozilla, "Link prefetching FAQ," https://developer.mozilla.org/en/Link_prefetching_FAQ.
[22] W3C, "HTML5 Link type prefetch," http://www.w3.org/TR/html5/links.html#link-type-prefetch.
[23] Google, "Web Developer's Guide to Prerendering in Chrome ": http://code.google.com/chrome/whitepapers/prerender.html.
[24] Google, "Announcing Instant Pages," http://googlewebmastercentral.blogspot.com/2011/06/announcing-instant-pages.html.
[25] J. Mickens, J. Elson, J. Howell, and J. Lorch, "Crom: Faster web browsing using speculative execution," in *Proceedings of the 7th USENIX conference on Networked systems design and implementation*, 2010.
[26] M. Rabinovich and O. Spatscheck, *Web Caching and Replication*, 2003.
[27] Z. Wang, F. X. Lin, L. Zhong, and M. Chishtie, "How effective is mobile browser cache?," in *Proceedings of the 3rd ACM workshop on Wireless of the students, by the students, for the students (S3)*, 2011.
[28] J. C. Mogul, "Hinted caching in the web," in *Proceedings of the 7th workshop on ACM SIGOPS European workshop: Systems support for worldwide applications*, 1996.
[29] WebKit, "The WebKit Open Source Project," http://webkit.org/.
[30] CyanogenMod Wiki, "Samsung Galaxy S II," http://wiki.cyanogenmod.com/wiki/Samsung_Galaxy_S_II.